# Estimate of convection-diffusion coefficients from modulated perturbative experiments as an inverse problem

F. Sattin[1], D.F. Escande[2], Y. Camenen[2], A.T. Salmi[3], T. Tala[3] and JET EFDA Contributors*

*JET-EFDA, Culham Science Centre, Abingdon, OX14 3DB, UK*

*[1]Consorzio RFX,Associazione EURATOM-ENEA, Padova, Italy*

*[2] UMR 7345, CNRS/Aix-Marseille Université, Marseille, France*

*[3] Association Euratom-Tekes, VTT, P.O. Box 1000, FI-02044 VTT, Finland*

## Abstract

The estimate of coefficients of the Convection-Diffusion Equation (CDE) from experimental measurements belongs in the category of inverse problems, which are known to come with issues of ill-conditioning or singularity. Here we concentrate on a particular class that can be reduced to a linear algebraic problem, with explicit solution. Ill-conditioning of the problem corresponds to the vanishing of one eigenvalue of the matrix to be inverted. The comparison with algorithms based upon matching experimental data against numerical integration of the CDE sheds light on the accuracy of the parameter estimation procedures, and suggests a path for a more precise assessment of the profiles and of the related uncertainty. Several instances of the implementation of the algorithm to real data are presented.

## 1. Introduction

Inverse problems are ubiquitous in applied sciences and engineering. They arise whenever one needs to extract from measurements some information about the object or the system measured. Their study represents almost a separate discipline, extending into sophisticated applied mathematics, with specialized scientific journals. Tomographic reconstruction represents the best known example; inverse scattering theory, both quantum and classical (e.g,

---

* See the Appendix of F. Romanelli *et al.*, Proceedings of the 23rd IAEA Fusion Energy Conference 2010, Daejeon, Korea

in acoustics, hydrodynamics or electromagnetism) is another one. A third instance, of paramount importance in magnetically confined plasmas--but relevant also in space physics (see, e.g., [1])--is the reconstruction of the equilibrium magnetic field by external magnetic field measurements *via* the Grad-Shafranov equation.

Transport Theory is another vast field of study: it encompasses the statistical description of the dynamics of some objects while moving through a host medium. Within this field, inverse problems are common, too: e.g., heat conduction [2]. Analysis can also be carried out at a more formal and abstract level, partially abstracting from the physical problems, and rather focussing on the mathematical structure of the partial differential equations that describe transport [3].

Inverse problems are notoriously often *ill-conditioned* or even *singular*, i.e., small variations in the data (due, e.g., to finite instrumental precision) yield extremely large (infinite) admissible ranges for the parameters to be estimated. Qualitatively: one single effect can hardly ever be unambiguously related to a single cause; rather several possibly widely different causes can be invoked. Quite often, inversion problems may reduce to linear algebraic ones. i.e., to inverting some matrices. In those cases, ill-posedness or singularity of the inverse problem amounts to the existence of small or null eigenvalues of the matrix to be inverted [4]. In this work we consider a simple but fairly important case in transport theory, namely the extraction of transport coefficients in one-dimensional convective-diffusive transport problems when the forcing term is periodically modulated. In principle, the mathematical formalism is quite generic and can be applied to a large variety of conditions; however, in this work it will be biased towards magnetic confinement plasma physics, where transport analysis is fundamental, either for comparison against predictions from fundamental theories, or for using it in extrapolating current scenarios, in the absence of a satisfactory theoretical basis. The algorithm itself is rather simple, thus is likely to have been rediscovered several times. In the context of fusion plasma physics, the first explicit solution to the problem was provided in 1990 by Krieger *et al* [5] within the framework of impurity transport, eight years later by Takenaga *et al* [6] in connection with transport of the main gas, and again this year by two of us, being focussed on heat transport. In this work we review the version of the solution as developed in [7]; for our purposes, this latter formulation has two advantages: (I) it is written in more compact, matricial form, which allows inspecting more easily its mathematical structure, and (II) takes explicitly into account the source; we will show later that it is actually the source to cause the problem to be locally singular. A large part of the discussion will be then devoted to inspecting to what extent the ill-posedness of the problem affects its solution.

## 2. Assessing convective-diffusive transport as an inverse problem

Convection-Diffusion equations arise in the contexts that involve the transport of one or more quantities through disordered media [8]. In the simplest version, only one passive scalar quantity ξ is considered in a one-dimensional geometry, the transport equation taking the form

$$\frac{\partial \xi}{\partial t} = -\nabla \cdot \Gamma(\xi) + S_\xi, \quad \xi = \xi(r,t) \quad \Gamma = -D\nabla\xi + V\xi \tag{1}$$

In (1), $S_\xi$ is the source/sink term, *r* the spatial coordinate.

In any experiment designed to measure some kind of transport, the experimentalist uses $S_\xi$ as a knob to vary ξ, which is the output of the experiment, measured on a spatial grid and with some time resolution: $\xi(r,t) \to \xi_{meas}(r_i,t_j)$. Inferring the transport coefficients $D, V$ amounts to finding expressions for them that, once inserted into Eq. (1), allow to extract solutions ξ matching $\xi_{meas}(r_i,t_j)$. Here lies the nature of *inverse problem*. It is trivial to show that, if we neglect time derivatives in (1), for a given measured profile $\xi_{meas}$, any couple ($D_0$, $V_0$) such that $\nabla \cdot \left(-D_0 \nabla \xi_{meas} + V_0 \xi_{meas}\right) = 0$ may be arbitrarily added to (*D*,*V*). Degeneracy of the solutions can be regarded as an extreme form of ill-conditioning: even perfect knowledge of the input is not enough to fully constrain the solution. Experiments are thus designed including finite time derivatives, in the belief that this extra information is able to remove the degeneracy. As we shall show, even explicit time dependence does not warrant curing the problem. To the best of our knowledge, as far as plasma physics is concerned, the first paper that explicitly takes into account the inverse nature of the problem and addresses the issue of ill-conditioning is [9]. Here, we anticipate the claim of the next section, i.e., that the problem is ill-conditioned. We consider a one-dimensional cylindrical geometry. Eq. (1), after rearrangement and a first integration over the radial coordinate, yields an expression for Γ

$$-D\frac{\partial \xi}{\partial r} + V\xi = -\frac{1}{r}\int_0^r dz\, z\left[S_\xi - \dot{\xi}\right] \tag{2}$$

where we use the zero-flux boundary condition at *r* = 0. If the source is time-dependent, this single equation is equivalent to infinitely many algebraic equations parameterized by the time $t$, once we consider $\xi$ , $\dot{\xi}$, $\partial_r\xi$ , $S_\xi$ , as known. Obviously, the system is clearly over determined if ($D$(r),$V(r)$) are the unknowns: this is just a manifestation of ill-conditioning. Hence, adding a time-dependence to the problem (1) alleviates but not completely removes the issues related to its invertibility.

## 3. A case admitting an explicit solution for (*D,V*)

Experiments are usually performed by adding a small perturbation to an initial equilibrium condition, hence $\xi$ in Eq. (1) must be understood as the difference between the instantaneous total quantity measured and its equilibrium value, and the same is true for $S_\xi$ . In any nonlinear system, the quantities in Eq. (1) are likely to depend from the starting equilibrium point; however, provided that the dependence is smooth and the perturbation weak enough, one can consider a linearization around the equilibrium that, hence, enters just as a parameter. We will postulate that this scenario holds; Eq. (1) becomes thus linear in *all* of the quantities appearing inside it.

As a fundamental simplification we consider those experiments where the source term is periodically modulated in time. This is not too a severe constraint, since most signals can be written as a finite sum of Fourier terms, and the algorithm is able to handle the case of either just one or a small finite number of harmonics. Furthermore, in magnetically confined plasmas, several classes of experiments, notably those of heat transport, are carried out *via* the periodic modulation of the source (say, RadioFrequency or Neutral Beam Injection Heating).

Taking advantage of its linearity, we can Fourier-transform Eq. (1) in the frequency domain

$$-i\omega\xi = -\frac{1}{\Im'(r)}\frac{\partial}{\partial r}\left[\Im'(r)\left(-D\frac{\partial \xi}{\partial r} + V\xi\right)\right] + S_\xi \qquad (3)$$

All the quantities appearing in Eq. (3), except for ($D,V$), are now understood to be complex numbers. The radial coordinate $r$ may be the true physical radius or any generalized coordinate with dimension of a length, and $\Im(r)$ is the volume enclosed inside $r$: for a cylindrical system, $\Im'(r) \propto r$, but generically $\Im$ can be a more complicated function of $r$ which, furthermore, may vary in time. As is convenient in this kind of experiments, the signal $\xi$ is written in terms of an amplitude and a phase: $\xi = Ae^{i\varphi}$, and Eq. (3) rewrites as a couple of real-valued equations.

By integrating once over the radius, and imposing the natural boundary condition

$\Gamma(r=0)=0$ at the center, we obtain two algebraic equations with unknowns (*D(r),V(r)*). They take a particularly convenient expression when expressed in matrix-vector form:

$$\mathbf{M}\cdot\mathbf{Y}=\mathbf{\Gamma} \tag{4}$$

where

$$\mathbf{Y}=\begin{pmatrix}D\\V\end{pmatrix},$$

$$\mathbf{M}=\begin{bmatrix}-A'\cos\varphi+A\varphi'\sin\varphi & A\cos\varphi\\ -A'\sin\varphi-A\varphi'\cos\varphi & A\sin\varphi\end{bmatrix}, \tag{5}$$

$$\mathbf{\Gamma}=\begin{pmatrix}\Im'(r)^{-1}\int_0^r\Im'(z)\left(\mathrm{Re}\left(S_\xi(z)\right)-\omega A(z)\sin\varphi(z)\right)dz\\ \Im'(r)^{-1}\int_0^r\Im'(z)\left(\mathrm{Im}\left(S_\xi(z)\right)+\omega A(z)\cos\varphi(z)\right)dz\end{pmatrix},\qquad f'\equiv\partial_r f$$

Eq. (4) can be inverted at each point *r*, to give the local values for $\mathbf{Y}(r)=(D(r),V(r))$. This formally solves the problem of estimating the transport coefficients, for the specific experiment considered.

Analysis of Eq. (4) leads to deep insight about the issues involved in the inversion problem. Actually, Eq (4) may be inverted whenever $\det(\mathbf{M})=A^2\varphi'\neq 0$. Since *A* must be nonzero in order to have a detectable signal, the inverse problem becomes ill-conditioned close to and at the singular points $r_s:\varphi'(r_s)=0$. Inspection of the literature shows that this condition is often met in experiments, and actually coincides always with the location of the source [10-15]. This is easily heuristically explained by noting that the l.h.s. of Eq. (3) comes as a balance between the two contributions of the r.h.s: the transport (Γ) and the source (*S*). If we suppose that when the source term is maximum, it dominates over the transport contribution, then we get $\xi\propto S$ and $S'=0\rightarrow\xi'=0$.

At any point *r* we can compute the two eigenvalues ($\lambda_{0,1}$) and eigenvectors ($\mathbf{E}_{0,1}$) of the matrix **M**: $\mathbf{M}\cdot\mathbf{E}_i=\lambda_i\mathbf{E}_i$. By writing $\mathbf{Y}=y_0\mathbf{E}_0+y_1\mathbf{E}_1,\mathbf{\Gamma}=g_0\mathbf{E}_0+g_1\mathbf{E}_1$, Eq. (4) becomes an equation for the two unknowns $y_{0,1}$. The eigenvalues and eigenvectors depend upon the data ($A$,$\varphi$): for instance, the explicit expression for λ's is

$$\lambda=\frac{1}{2}\left[-A'\cos\varphi+A(1+\varphi')\sin\varphi\pm\sqrt{\left(A'\cos\varphi-A(1+\varphi')\sin\varphi\right)^2-4A^2\varphi'}\right] \tag{6}$$

At the singular points, one of the eigenvalues vanishes: $\lambda_0(r_s)=0$. Accordingly, the l.h.s. of Eq. (4) aligns along the other eigenvector

$$\mathbf{M}\cdot\mathbf{Y}=\mathbf{M}\cdot(y_0\mathbf{E}_0+y_1\mathbf{E}_1)=\lambda_1 y_1\mathbf{E}_1=\mathbf{\Gamma},\qquad r=r_s \tag{7}$$

Physically, transport coefficients must be defined everywhere, hence Eq. (7) must admit a solution at $r = r_s$. This is possible only if its r.h.s. aligns along $\mathbf{E}_1$, too: $\mathbf{\Gamma}(r_s) = g_1\mathbf{E}_1$. Unavoidable errors present in any measurement make unlikely this case, i.e., using the data from the experiment, one expects to find points $r_s$ where $\lambda_0(r_s) = 0$ but $\mathbf{\Gamma}(r_s) = g_0\mathbf{E}_0 + g_1\mathbf{E}_1$, with $g_0(r_s) \neq 0$. More generally, any perturbation of the values λ, **E**, reflects upon **Y** as

$$\mathbf{Y} = \frac{g_0}{\lambda_0}\mathbf{E}_0 + \frac{g_1}{\lambda_1}\mathbf{E}_1 \rightarrow \delta\mathbf{Y} = \left(\frac{\delta g_0}{\lambda_0} - \frac{g_0}{\lambda_0^2}\delta\lambda_0\right)\mathbf{E}_0 + \left(\frac{\delta g_1}{\lambda_1} - \frac{g_1}{\lambda_1^2}\delta\lambda_1\right)\mathbf{E}_1 + \frac{g_0}{\lambda_0}\delta\mathbf{E}_0 + \frac{g_1}{\lambda_1}\delta\mathbf{E}_1 \quad (8)$$

Eq. (8) may be used to estimate visually how given errors on the raw data, translated into error bars for λ, **E**, propagate onto **Y**. It is apparent how δ**Y** blows up when the singular points are approached:

$$\delta\mathbf{Y} \approx -\frac{g_0\delta\lambda_0}{\lambda_0^2}\mathbf{E}_0, \lambda_0 \rightarrow 0 \quad (9)$$

Even if $\mathbf{\Gamma}(r_s) \propto \mathbf{E}_1$ and Eq. (7) becomes a well-defined equation, it does not place any condition upon the full solution **Y** at $r = r_s$, but only on a subspace of it: i.e., it fixes $y_1 = g_1/\lambda_1$ whereas the component along the subspace parallel to $\mathbf{E}_0$ is undefined: any arbitrary vector aligned along the local $\mathbf{E}_0$ eigenvector can be added to the solution and still remain compatible with the data. This is where the degeneracy of the problem enters in. An example is shown in Fig. (1). The clouds in the two contour plots stand for the typical distributions of the couples ($D$,$V$) that one expects after a Monte Carlo simulation taking into account errors on the data. The thick black lines are parallel to the local $\mathbf{E}_0$ eigenvector. The left plot is for a regular point: independent estimates distribute roughly as a normal distribution around the average value. The right plot shows what happens when a singular point is approached: the distribution aligns along $\mathbf{E}_0$.

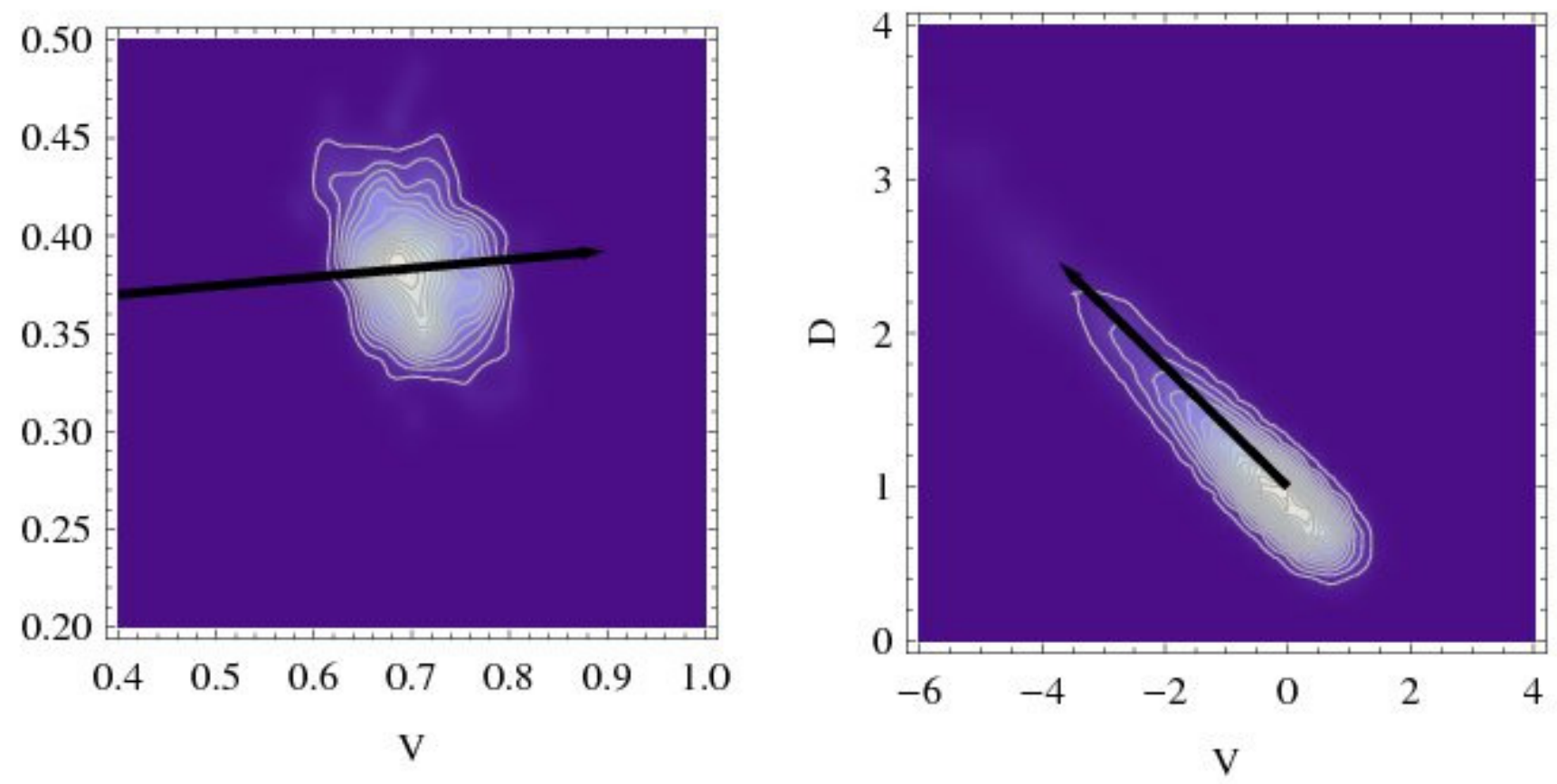

**Fig. 1.** Contour plot for the statistical distribution of (*D*,*V*) couples. Data are produced by 400 independent realizations of synthetic data with stochastic perturbations added. Black lines are parallel to the local $\mathbf{E}_0$ eigenvector. On the left, a regular point; on the right a singular point.

If we knew *exactly* **Y** at all regular points $r \neq r_s$, then $y_0(r_s)$ could easily be fixed by imposing continuity of the solution there. However, this is not the case: Eq. (8) tells us that the unavoidable presence of errors in the measurement of ($A$, φ) implies some ignorance upon $y_{0,1}$ even at the regular points. It may be extremely small far from the singular points, but unavoidably increases without bound as soon as we approach $r_s$ (Eq. 9); hence, for practical purposes we cannot have an estimate for $y_0$ not only at $r_s$ but also near to it.

## 4. On alternative approaches to estimating (*D*,*V*) and the issue of regularization

Inferring from experiment transport coefficients (*D*,*V*) is an important issue in plasma physics, since the complexity of plasmas makes first-principles calculations quite hard[1]. Accordingly, some ingenuity has been exerted on the numerical side. Paper [16] presents a recent and extensive review of the state of the art of perturbative transport experiments, including details of numerical algorithms and related issues. The most adopted recipe (see, e.g., instances in [10-15]) makes use of transport codes[2], i.e., numerical codes that, for a given geometry and *for a given transport*, compute the resulting profile ξ. The root-finding procedure is thus no longer direct, but rather iterative and can be split into the following steps: (*I*) First, an analytical expression is chosen for (*D,V*), containing a number of adjustable parameters $\{\alpha\}$: Let us label these expressions with $D_{\{\alpha\}}$, $V_{\{\alpha\}}$. (*II*) A first guess of ($D_{\{\alpha\}}$, $V_{\{\alpha\}}$) is used to integrate Eq. (1) or Eq. (3), yielding as output a simulated profile for $\xi$: $\xi_{sim}$. (*III*) The simulated profile is compared against the measured one, $\xi_m$, and the difference $\Delta = \xi_{sim} - \xi_m$ is computed. (*IV*) The steps (*II,III*) are iterated, varying the adjustable parameters $\{\alpha\}$ as long as $\Delta$ is minimized under some given global norm.

---

[1] In view of the difficulty in carefully assessing (*D*,*V*), sometimes simplified approaches are used, where it is *postulated* that the whole transport may be accounted by just one mechanism, the diffusion, thus requiring just *D* to be measured. In earlier works [17,18] we showed that it is not always justified.

[2] Of course, this same procedure may hold for other fields, not just plasma physics.

The above steps implicitly postulate that, for a suitable choice of $\{\alpha\}$, ($D_{\{\alpha\}}$, $V_{\{\alpha\}}$) can always collapse to the true value ($D$,$V$). We have shown that this is not warranted. At a singular point, the coefficients ($D$,$V$) that solve Eq. (3) for some set of data ($A$,$\varphi$) must fulfil Eq. (7) which, as we stated earlier, provides a constraint upon only the subspace $\mathbf{E}_1$ of the solution. We demonstrated rigorously that the component aligned along $\mathbf{E}_0$ *cannot* be determined by the Eq. (3) alone: Eq. (9) shows that $\delta\mathbf{Y}(r_s)$ spans the whole axis $\mathbf{E}_0$ regardless of the magnitude of the experimental errors upon the data ($A$,$\varphi$). Nonetheless, any numerical minimization procedure used in step (*IV*) will unavoidably converge towards some set $\{\alpha\}$, thereby yielding an estimate for the *full* solution, along with *finite* estimates for the errors: $\{\delta\alpha\}$. This happens since codes perform a global minimization: the couple ($D_{\{\alpha\}}$, $V_{\{\alpha\}}$) at a point $r$ is determined by a minimization that involves data at all points, and is furthermore constrained by the set of basis functions chosen to parameterize the transport coefficients. This is an implicit *regularization* of the solution. Regularization is an essential ingredient in ill-posed problems. It amounts to adding supplementary constraints to the solutions in order to drop any undesirable feature (e.g., lack of smoothness, excessive oscillations,…). In Bayesian jargon: impose prior distributions on model parameters. These constraints *by definition* may be not dictated by the physics of the problem at hand, and are often based upon subjective considerations; therefore, *a priori* there is not warranty that they are the most appropriate. In this specific case, the regularization imposed by the code is not harmful far from the singular points, but close to them its validity needs careful examination in order not to constrain the solution between artificially small error bars. We will meet one such an instance in the next section.

If we wish to apply Eq. (4) to experiments, we have to deal with measurements $\{A_{(l)}, \varphi_{(l)}\}$ taken at a discrete set of points $r = r(l)$. This leads to the issues of data interpolation and extrapolation. Interpolation between neighbouring points is necessary to compute the derivatives that appear in the matrix $\mathbf{M}$ as well as to compute the integrals in $\mathbf{\Gamma}$. Extrapolation enters in connection with $\mathbf{\Gamma}$, too, since its definition requires computing integrals from $r = 0$ onwards, whereas data are taken within some finite range, usually not including the origin. We will consider in detail these issues in section 6, in connection with analysis of true experimental data. Here, we highlight a difference between the present approach and the transport codes: Eq. (1) being a second-order differential equation, for solving it one needs two boundary conditions. The first one is the already mentioned zero-flux condition at the origin; as second condition, it is commonly taken the value of the field ξ at the outermost measurement location, with the consequence that (*D*,*V*) can no longer be estimated there. The present approach does not need this second boundary condition, and therefore saves for the modelling some (possibly valuable) information.

One rather surprising result appears when one looks how errors upon measurements[3] $\left(A_{(l)} \pm \delta A_{(l)}, \varphi_{(l)} \pm \delta\varphi_{(l)}\right)$ propagate onto **Y**. Let **W** = $\mathbf{M}^{-1}$. Writing (4) using Einstein's convention on indices yields

$$Y_i(r(l)) = W_{ij}\Gamma_j$$

$$\rightarrow \delta Y_i(r(l)) = \left(\frac{\partial W_{ij}}{\partial A_{(l)}} + \frac{\partial \Gamma_j}{\partial A_{(l)}}\right)\delta A_{(l)} + \left(\frac{\partial W_{ij}}{\partial \varphi_{(l)}} + \frac{\partial \Gamma_j}{\partial \varphi_{(l)}}\right)\delta\varphi_{(l)} + \sum_{k<l}\left(\frac{\partial \Gamma_j}{\partial A_{(k)}}\delta A_{(k)} + \frac{\partial \Gamma_j}{\partial \varphi_{(k)}}\delta\varphi_{(k)}\right) \quad (10)$$

The total error upon $Y_i$ is due to a local contribution (depending upon *r(l)* alone) that includes the first two terms, and another one from the data taken at points *r(k)* < *r(l),* which appear because of their presence within the integrals. Allowing for independently distributed errors, the sum grows in absolute value roughly like the square root of the number of summands (i.e., roughly as $r(l)^{1/2}$). Hence, Eq. (10) hints to a generic propensity of modelling to become less and less accurate as the radius grows, although the precise behaviour can easily be reversed in particular cases due to the specific dependence of **W** and **Γ** upon $\left\{A_{(l)}, \varphi_{(l)}\right\}$,

## 5. Multiple experiments and extended transport models

The algorithm in section 3 refers to a very specific case, where a single Fourier component from *one* experiment is taken. However, nothing prevents considering simultaneously several experiments or Fourier modes. The only unavoidable constraint is that the plasma conditions remain the same, so that one can reasonably claim that transport (i.e., *D,V*) is the same, too. Each set of data yields an instance of Eq. (4): $\mathbf{M}_i \cdot \mathbf{Y} = \mathbf{\Gamma}_i$ . This leads to an overdetermined problem, that allows recovering a common estimate for **Y** via, e.g., some weighted average.

---

[3] In actual experiments, the source *S*, too, is only known to within some finite precision. However, including errors on it in the analysis is straightforward.

Let us consider the simplest case of two data sets . At any point $r$ we get two estimates, each with its error bar, $\mathbf{Y}_1 \pm \Delta\mathbf{Y}_1, \mathbf{Y}_2 \pm \Delta\mathbf{Y}_2$. If the two values are consistent, within the error bars, we can extract a common estimate endowed with an error bar wich is given by the overlap between the two independent estimates, hence is smaller than both $\Delta\mathbf{Y}_1$ and $\Delta\mathbf{Y}_2$. This is extremely valuable when $r$ is a singular point for, say, $\mathbf{Y}_1$ , since we can estimate the local $\mathbf{Y}$ using $\mathbf{Y}_2$ . A different case is when, at all points, $\mathbf{Y}_1 \approx \mathbf{Y}_2, \Delta\mathbf{Y}_1 \approx \Delta\mathbf{Y}_2$. This means that we are practically dealing with two repetitions of the same experiment, hence no new information can be gained this way. The opposite and interesting case is when, even taking into account the error bars, the two estimates cannot agree: $\mathbf{Y}_1 \neq \mathbf{Y}_2$. This conflicting evidence is the proof that something is wrong either at the level of the data (i.e., at least one set of measurements is flawed), or at the level of modelling (that is, Eq. (1) is not the appropriate framework for modelling this kind of transport, or the transport coefficients are not the same between the two experiments).

The redundancy provided by multiple experiments can also be employed for assigning values to further parameters, within the framework of models that extend the physics beyond the convective-diffusive picture of Eq. (1). This is the case, e.g., for the transport of toroidal angular momentum in magnetically confined plasmas. As explained in detail in [19], an accurate modelling of toroidal rotation Ω might require the explicit consideration of additive contributions to the flux which do not directly depend upon the rotation nor to the external torque. At this stage, we do not afford a full-fledged investigation attempting to identify separate ingredients to this contributions: in principle, they could be Neoclassical Toroidal Viscosity, residual stress terms, some ignorance about the NBI modulation, or other unidentified torques. In this work, we limit to the simple question: is it possible from the data and using the above method to infer anything about the possible existence of additional flux terms in Eq. (3), that do not directly depend upon the measured quantity (toroidal momentum in our case)? An additional contribution to the flux writes, for the equilibrium and modulation:

$$\Gamma = -D\nabla\Omega + V\Omega \rightarrow -D\nabla\Omega + V\Omega + R \quad (11)$$

and, for perturbed quantities:

$$\Gamma \rightarrow \Gamma + \delta\Gamma, \quad \delta\Gamma = -D\nabla\delta\Omega + V\delta\Omega + \delta R \quad (12)$$

After temporal Fourier transform, we may thus write the like of Eqns. (4-6):

$$\mathbf{M} \cdot \mathbf{Y} + \mathbf{R} = \mathbf{\Gamma} \quad (13)$$

where $\mathbf{R}$ is an array containing the real and imaginary part of $R$. At this level of description, it must be treated as an unknown, on the same footing as **Y.** Hence, Eq. (13) alone is not sufficient for fixing all the variables. However, using two independent experiments, we can:

$$\begin{cases} \mathbf{M}_1 \cdot \mathbf{Y} + \mathbf{R} = \mathbf{\Gamma}_1 \\ \mathbf{M}_2 \cdot \mathbf{Y} + \mathbf{R} = \mathbf{\Gamma}_2 \end{cases} \tag{14}$$

It is now straightforward to solve Eqns. (14) for (**Y**,**R**).

To conclude this section, we recall that an important and straightforward instance of multiple-experiment analysis is by considering, together with modulation data, also the steady-state profiles. This information is often used in transport codes as a useful constraint.

## 6. Simulations of experimental data

We present two instances of the usage of the present algorithm, all of them based upon JET data. Let us summarize in a brief recipe how calculations are performed: (*a*) the fundamental bricks are the data $\{A_{(l)}, \varphi_{(l)}\}$ along with the source *S*. Beforehand of the analysis they are interpolated using smooth curves (cubic splines proved to work fine). However, if the data look noisy, this procedure alone is not sufficient. The reason is that random fluctuations between neighbouring points, particularly if the average slope is small, may yield interpolating curves whose derivatives vanish at several points. These singular points are actually artefacts, but in view of the analysis of Section 3, they produce wild fluctuations in the local estimate of transport coefficients. As a remedy, in the case of noisy data, we preprocessed them with a Gaussian filter. (*b*) The functions *A*(r), φ(r), *S*(r) are used to compute the entries of the matrix **M** as well as of **Γ**. The remaining part of the integrals, from $r = 0$ to $r_{(1)}$, the innermost measurement point, have been approximated using the trapezoidal rule: $\int_0^{r_{(1)}} z\, f(z)\, dz \approx (r_{(1)})^2 f(r_{(1)})/2$. (*c*) The couple (*D*,*V*) at each point is recovered from inversion of Eq. (4) [or, the triple (*D*,*V*,*R*) from the system (14)]. (*d*) The sensitivity study is carried out via Monte Carlo techniques: each datum is at a first stage independently varied $(A_{(l)}, \varphi_{(l)}) \rightarrow (A_{(l)} \pm \delta A_{(l)}, \varphi_{(l)} \pm \delta\varphi_{(l)})$, where $(\delta A_{(l)}, \delta\varphi_{(l)})$ are randomly picked up from a normal distribution with zero mean and standard deviation given by experimental errors. The steps (*a-c*) are then repeated with the new set of data, and new coefficients (*D*,*V*) computed. Repeating the whole procedure over a large number of independent runs yields a statistical distribution for (*D*,*V*), from which confidence intervals can be drawn.

The first dataset refers to JET pulses 73701-2,73704,73707-9. They are part of a set of discharges explicitly designed for measuring momentum transport, as documented in the paper [20]; hence, represent an excellent workbench for the present method. Here we do not provide informations about the experiments' setup, referring the reader to the paper [20] . Only the relevant output is shown In Fig. (2): the amplitude and phase of the measured signal (the toroidal velocity of rotation), taken both at the fundamental frequency ($\nu_1$ = 8.33 Hz) and the second harmonic ($\nu_2 = 2\times\nu_1$), together with the radial profile of the torque (obtained by modulating the NBI). Fig. (3) presents the corresponding ($D$,$V$) couples computed for both harmonics using Eq. (4). Data have been prefiltered, with a kernel's width equal to 20% of the minor radius[4]. Error bars have been built by running 20 independent realization of the data, each point being perturbed with a Gaussian noise of amplitude 20% (in the amplitude $A$) or 7.5/15 degrees (in the phase $\varphi$, depending on the harmonic). Our results should be compared against, e.g., Figs. (9a,b) of ref. [20]. There is qualitative agreement, since both studies yield increasing trends for $D$ and $|V|$ in the outward direction, with a ratio of about (2 ÷3) between the edge and the core values, whereas our results exceed somewhat quantitatively of [20]. If we compare the results obtained by analizing the data relative to the two frequencies $\nu_1$, $\nu_2$ , they are fully compatible, although the fundamental harmonic yields more accurate estimates, having the larger amplitude.

---

[4] The need for filtering small scales is apparent from Fig. (2): the fluctuations in the phase make several spurious singular points ($\varphi$ ' = 0) to appear, which make impossible reconstruct trasport coefficient there. However, it can also be justified on the basis of the way the data in fig. (2) were produced: they come from the mapping on plasma coordinates of measurements taken at 12 locations on the equatorial plane. This means that roughly any structure spanning less than about 1/10 of the radius is not likely to have physical meaning.

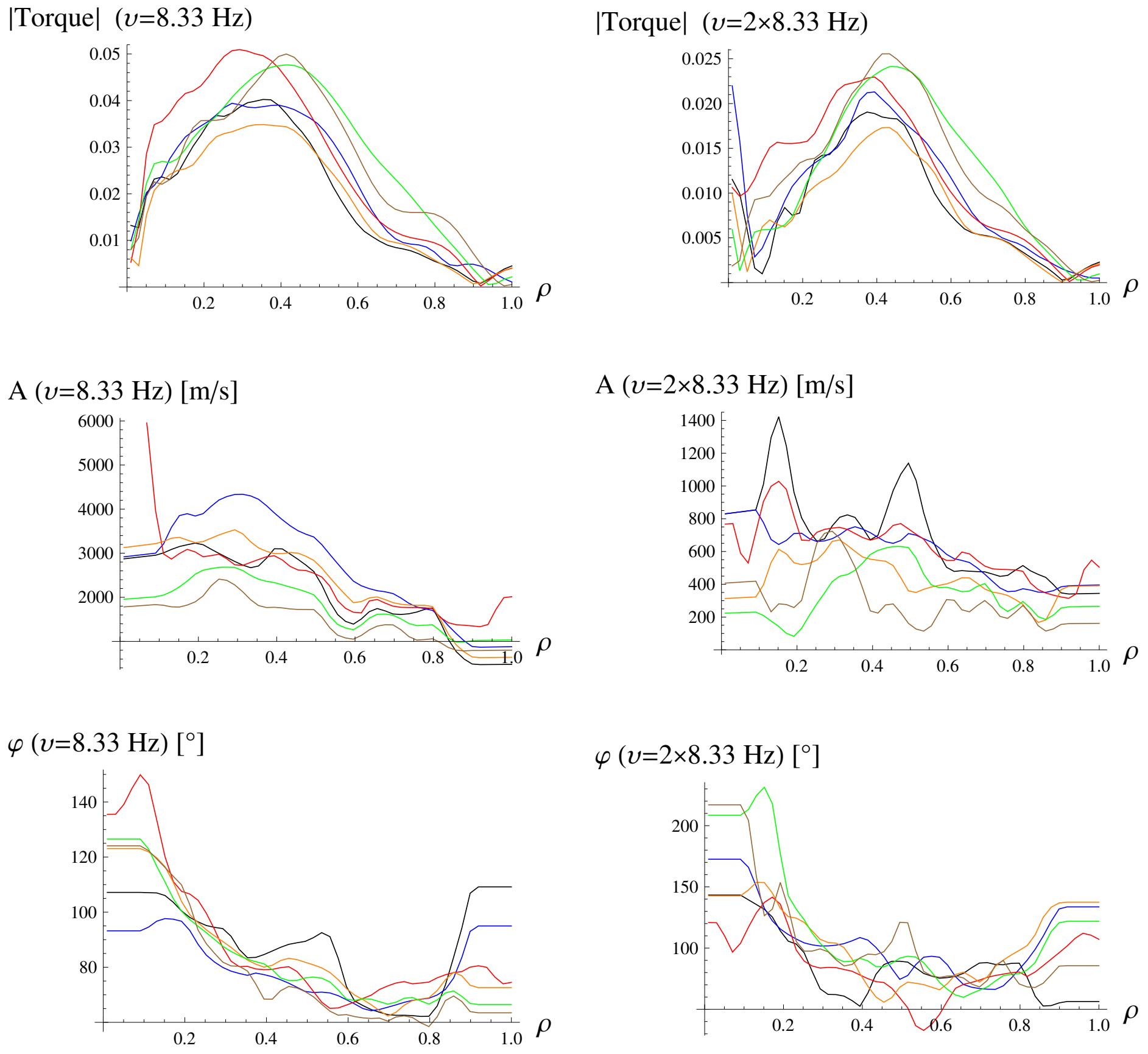

**Fig. 2** From top to bottom: torque density (N m$^{-2}$), amplitude $A$ of the perturbation, phase φ. In the left column there are the data for the fundamental frequency of modulation, in the right column those for the first overtone. The color code is: black, shot 73701; blue, 73702; red, 73704; orange, 73707; green, 73708; brown, 73709. Here and in all the other plots, the horizontal coordinate ρ is the square root of the normalized toroidal flux

There is some shot-to-shot difference in the ($D$,$V$) profiles. A more complete investigation is obviously desirable attempting to assess what is the cause. It is beyond the scope and the length limits of this work, and will be postponed to future studies. However, for the moment we note that the discharges are not identical: they differ slightly in the equilibria (density scale length, $q$ profile, and equilibrium toroidal rotation), hence it is quite likely that at least part of the differences be due to this reason.

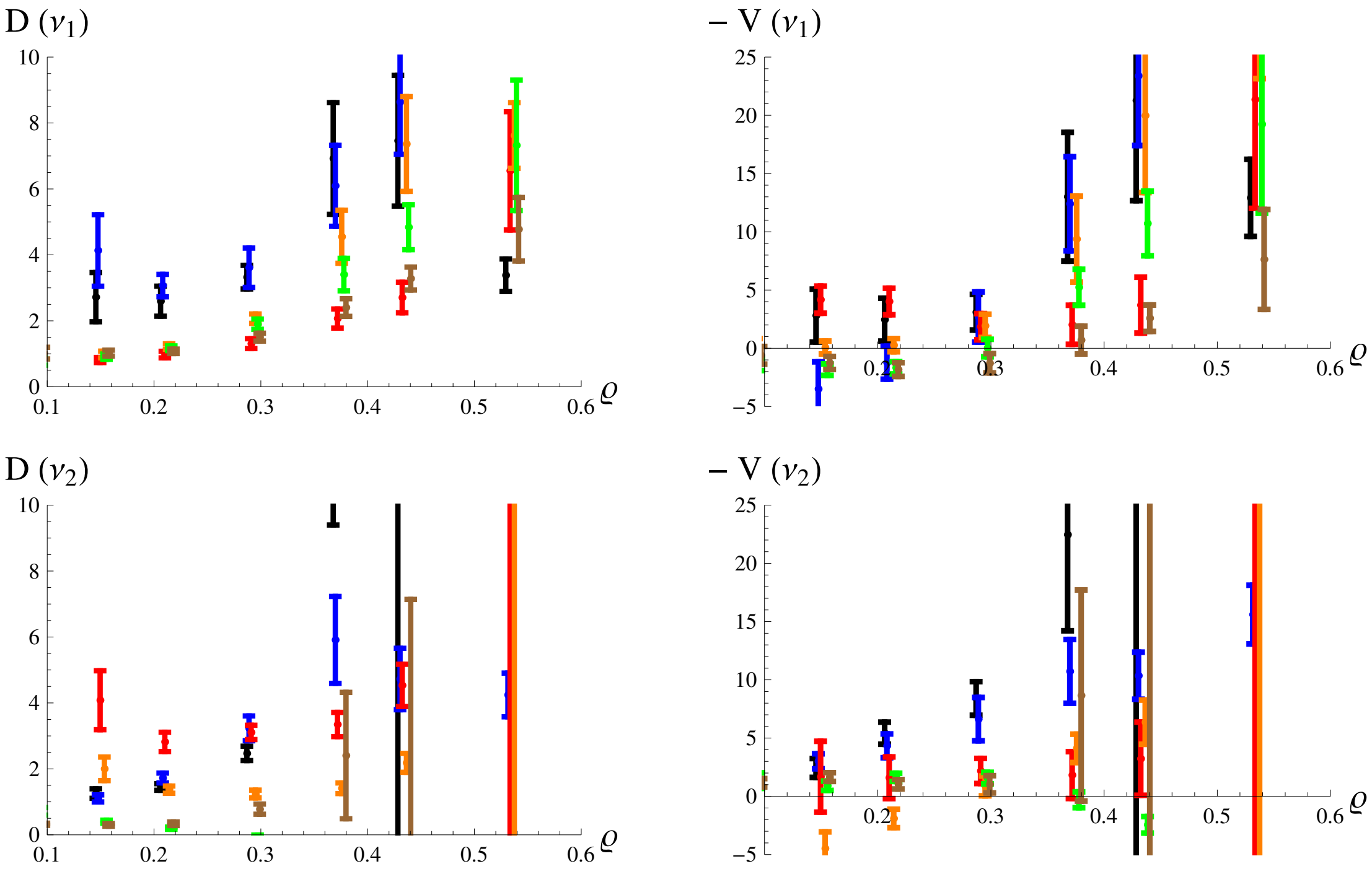

**Fig. 3** Transport coefficients computed from the data of Fig. (2): Diffusivity $D$ (m$^2$/s) and (- pinch V) (m/s) for both the fundamental frequency of modulation $\nu_1$ and first overtone $\nu_2$. The color code is the same as Fig. (2). Different sets of data have been slightly horizontally shifted with respect to each other. Reconstructed coefficients span the range $\rho < 0.55$ since in the region $0.6 < \rho < 0.8$ the phase of the signal is almost constant (fig. 2, bottom row), making the reconstruction unfeasible: this is made apparent by the large error bars at the largest radii. For the same reason, some points are missing below $\rho = 0.2$.

As a second dataset we refer to the shots extensively reported in the paper [15]: a set of JET L-mode-confinement discharges designed to study the transport of toroidal momentum in the presence of different torques and of magnetic ripple. The presence of finite ripple is expected to exert some supplementary torque, hence these discharges appear promising candidates for detecting further contributions to the flux, $R$. We considered in detail shots 77090, 77091: in these shots, the imposed ripple was quite similar, except for a small difference in the equilibrium rotation, hence, we could speculate that the additional contribution is almost the same and use the formalism of Eq. (14). Figure (4) shows the ($A$,φ, $S$) profiles (only the fundamental harmonic is considered in this analysis), and Fig. (5) the results ($D$,$V$,$R$); 20 independent statistical realizations were used for getting the error bars.

Notice that, from Fig. (5), the three contributions carry approximately equal fractions of the total flux: $D\partial_r\xi \approx V\xi \approx R$ .

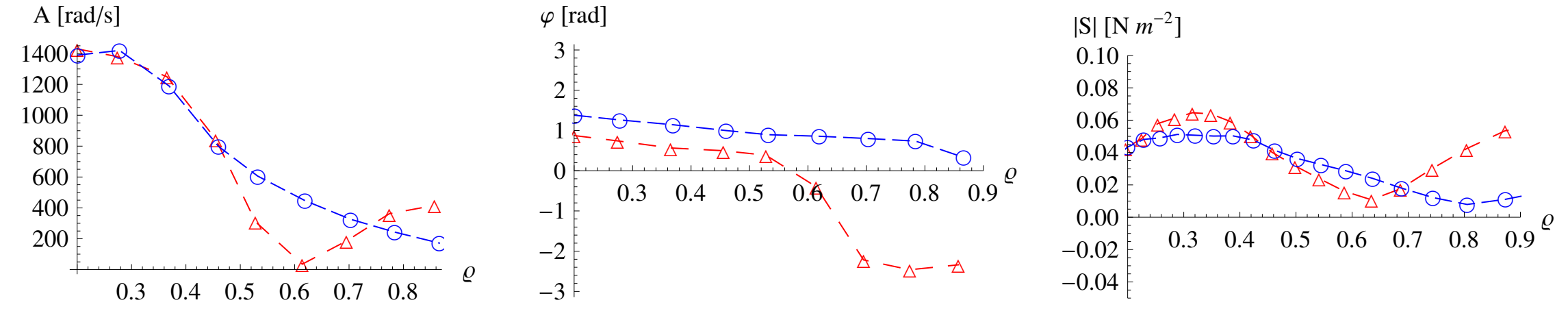

**Fig. 4** Left, amplitude; center, phase; right, torque. The data are the same published in ref. [20] and refer to the fundamental modulation frequency ν = 6.25 Hz. Red curves with triangles refer to shot 77090; blue curves with circles, to shot 77091.

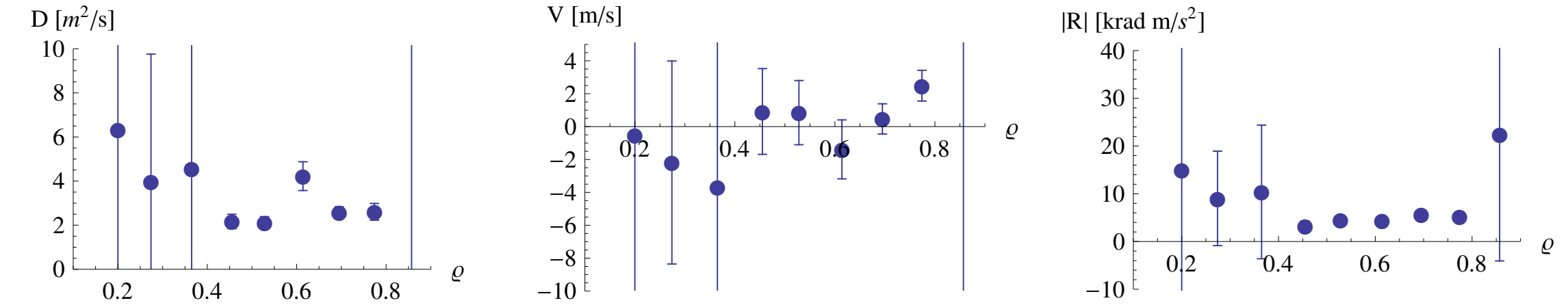

**Fig. 5** Left, diffusivity *D*; center, pinch *V*; additional flux, *R*, obtained from solving Eq. (14) with the data of Fig. (4). The results refer to the fundamental harmonic, ν = 6.25 Hz.

In a second simulation (Fig. 6), we compute the couples (*D*,*V*) separately from each shot without accounting for the additional term *R* appearing in Eqns (11-14): $R \to 0$. There is good agreement between the two shots for radii smaller than about 0.5. This suggests that, inside of this region, the presence or lack of *R* is not a discriminant. Conversely, in the outermost half radius, the two *D* estimates in Fig. (6) are clearly incompatible. This implies that either the two shots do not share the same set of transport coefficients, or that some contribution from *R* is needed. Since *D* in shot 77090 (red triangles) takes vanishingly small values—whose physical meaning is dubious—it is plausible to argue that a correct book-keeping can be restored only with the second option.

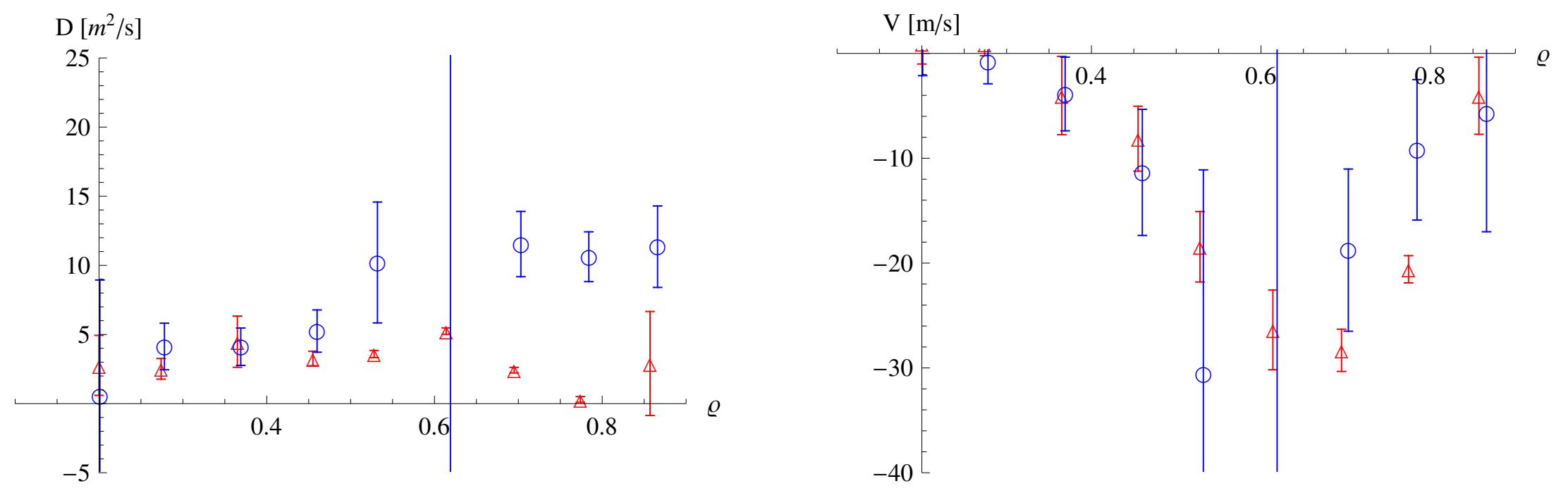

**Fig. 6** Left, diffusivity $D$; right, pinch $V$ for the shots 77090 (red symbols), and 77091 (blue symbols). Transport coefficients are computed using Eq. (4). i.e., neglecting $R$.

## 7. Conclusions

To summarize, we have discussed a computationally very light approach (hereafter, the Matricial Approach--MA) to the inversion of the one-dimensional Convective-Diffusive Equation (1) under periodic forcing. Its major advantage is that it is direct, providing an explicit solution. This avoids to a large extent all the issues typical of iterative methods, that include the choice of the parameterization of the class of trial solution functions—and more generally their regularization—and the numerically heavy minimization procedure. Furthermore, it provides a clear geometrical foundation to the nature and size of uncertainties in profile reconstruction. The reconstruction radius-by-radius enables to make explicit the local uncertainty $\delta\mathbf{Y}(r)$. The algebraic inversion yields a high precision in the reconstruction of transport profiles: indeed, this method is not restricted by the *a-priori* guess of the trial profiles, but by that of these derivatives of $(A,\varphi)$, which is generally much more reliable and controllable. In the presence of singular points, the huge uncertainty related makes the estimate of $\mathbf{Y}(r)$ useless for practical purposes: in this case, some regularization is still useful. This is achieved by overlapping the solutions' uncertainty intervals from various experiments where the same transport is assumed to hold: it provides either a way to improve the precision of the reconstruction (case of a non vanishing overlap) or to prove the set of initial assumptions in the reconstruction to be wrong (case of a vanishing overlap). The MA can help designing *a priori* the combination of perturbation measurements susceptible of improving the precision of the reconstruction of transport profiles. It can also be easily extended to include parameters other than $D$ and $V$.

## Acknowledgments

*This work was supported by EURATOM and carried out within the framework of the European Fusion Development Agreement. The views and opinions expressed herein do not necessarily reflect those of theEuropean Commission.The authors thank R. Paccagnella for careful reading of the manuscript.The authors wish to thank an anonymous referee for pointing ref. [5] to their attention and for other useful suggestions.*